\documentclass[letterpaper]{article}

\usepackage{natbib,alifeconf}  
\usepackage{url}
\usepackage{amsmath}
\usepackage{hyperref}
\usepackage{cleveref}
\usepackage{booktabs}

\usepackage{subcaption}
\title{ALIFE2024 template}

\title{A Simulation Environment for the Neuroevolution of\\Ant Colony Dynamics}

\author{
    Michael Crosscombe$^{*}$,
    Ilya Horiguchi,
    Norihiro Maruyama,
    Shigeto Dobata, \and
    Takashi Ikegami \\
    \mbox{}\\
    Department of General Systems Studies, Graduate School of Arts and Sciences, The University of Tokyo, Japan \\
    $^{*}$cross@sacral.c.u-tokyo.ac.jp
} 

\begin{document}

\maketitle

\begin{abstract}
    We introduce a simulation environment to facilitate research into emergent collective behaviour, with a focus on replicating the dynamics of ant colonies. By leveraging real-world data, the environment simulates a target ant trail that a controllable agent must learn to replicate, using sensory data observed by the target ant. This work aims to contribute to the neuroevolution of models for collective behaviour, focusing on evolving neural architectures that encode domain-specific behaviours in the network topology. By evolving models that can be modified and studied in a controlled environment, we can uncover the necessary conditions required for collective behaviours to emerge. We hope this environment will be useful to those studying the role of interactions in emergent behaviour within collective systems.
\end{abstract}

\section{Introduction}
\label{sec:intro}
In the pursuit of novel forms of Collective Intelligence (CI) living systems provide us with extensive datasets for understanding the fundamental properties of collective behaviour. That is, living systems are useful for studying how local interactions between individuals lead to the emergence of global behaviours in the population through a process of self-organisation~\citep{Couzin2003}. However, a full understanding of the necessary conditions for collective behaviour cannot come from merely observing living systems. Instead, we must study said conditions through controlled experiments that, while inspired by natural systems, allow for the evolution and manipulation of models and their variables to observe the effects.

Simulations afford the opportunity to model individual behaviours and interactions within a controlled setting, enabling the study of emergent properties in a systematic and repeatable manner. Recently, efforts to merge deep learning and complex systems has greatly improved CI research capabilities by enabling simulations to scale to hundreds or thousands of individuals; many more than was previously possible~\citep{Ha2022}.
By leveraging real-world data from living systems, we can simulate environments with the same conditions in which collective behaviours naturally occur, allowing for the exploration of factors such as communication, information aggregation~\citep{Kameda2022}, coarse-graining~\citep{Flack2017}, and their impacts on the emergence of collective behaviours.


Here we introduce a \href{https://github.com/Farama-Foundation/Gymnasium}{Gymnasium}~\citep{towers_gymnasium_2023} environment to facilitate research into the necessary conditions for emergent collective behaviour\footnote{\scriptsize \url{https://github.com/TooHuman/AntDynamics-Gymnasium}}.
Our environment presents a target ant trial (from real data) that the learning agent attempts to replicate, mimicking the dynamics based on relative sensory data observed by the target ant along the trail (e.g., nearby ants).

\section{Simulating Ant Dynamics}
\label{sec:data}

\begin{figure}[t]
    \begin{center}
        \begin{subfigure}{.22\textwidth}
            \includegraphics[width=1\textwidth]{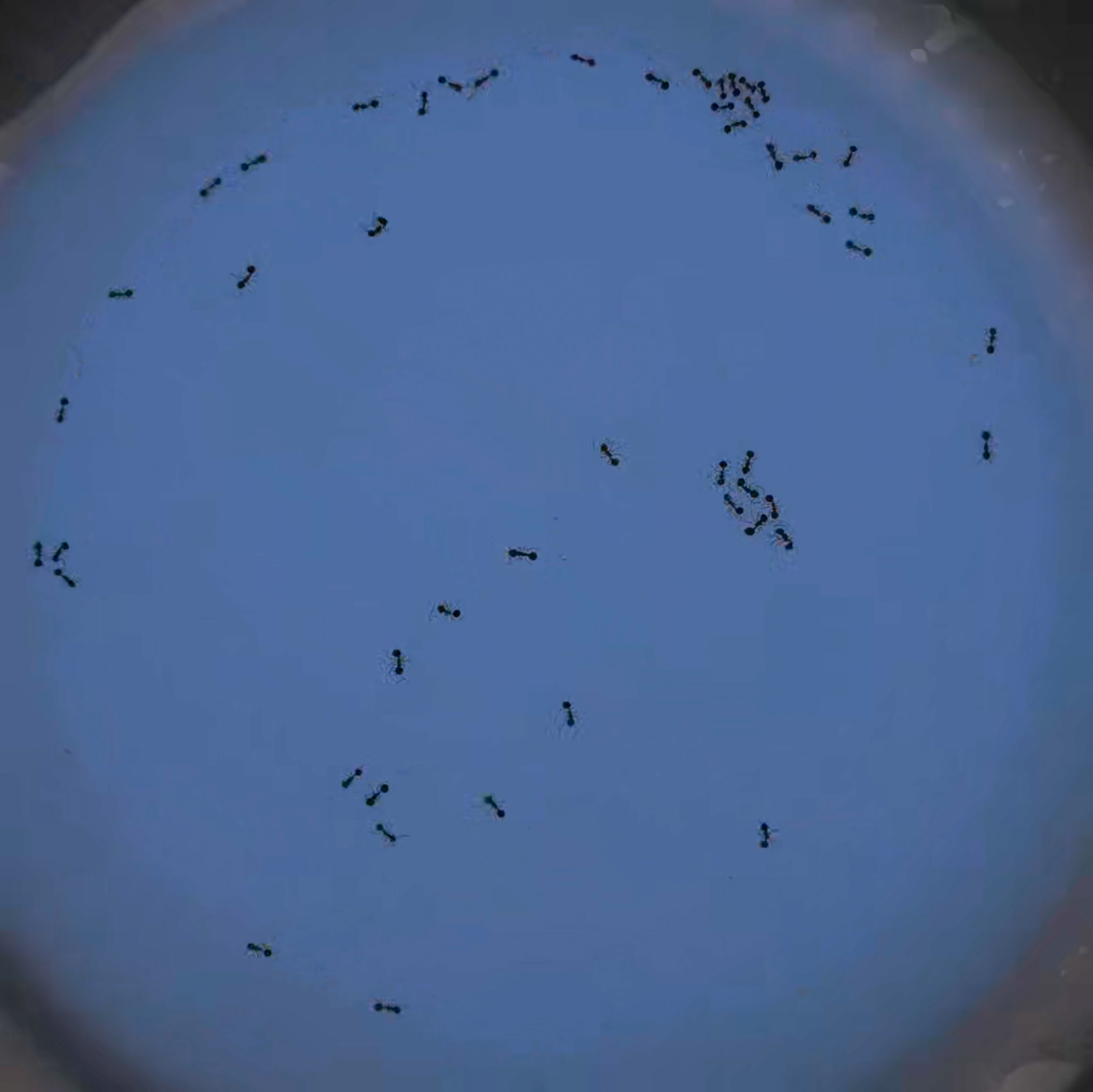}
        \end{subfigure}
        \begin{subfigure}{.22\textwidth}
            \includegraphics[width=1\textwidth]{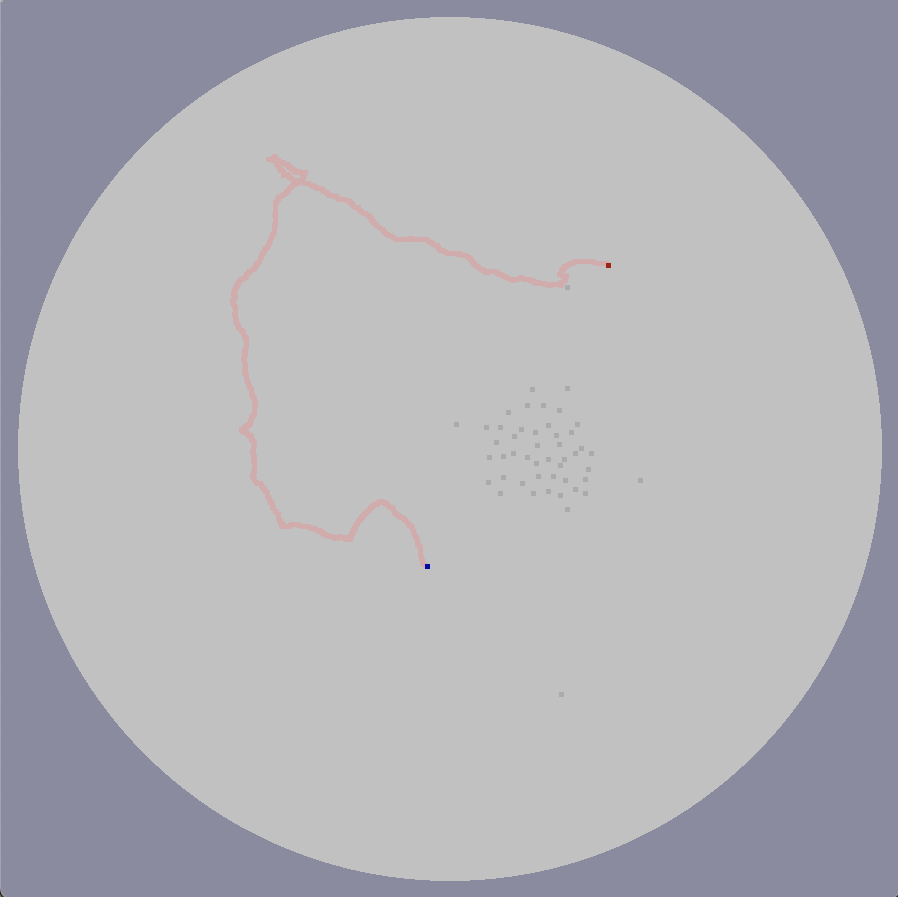}
        \end{subfigure}
        \caption{\label{diag:arena-sim}
        \textbf{Left:} Top-down video recordings, cropped and scaled to $1280$x$1280$ resolution. The ants are filmed from above in an evenly-lit $100$ mm diameter arena. Individual ant positions are extracted from a $4$ hour recording.
        \textbf{Right:} Our simulation environment using \href{https://www.pygame.org/wiki/about}{Pygame} and Gymnasium to reproduce ant colony dynamics with an agent that can interface with a policy network.}
    \end{center}
\end{figure}

\begin{figure}[t]
    \begin{center}
        \begin{subfigure}{.30\textwidth}
            \includegraphics[width=1\textwidth]{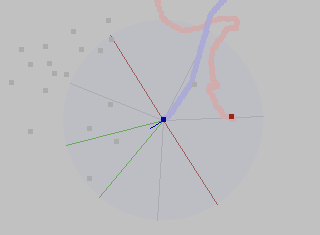}
        \end{subfigure}
        \caption{\label{diag:ant-vision}
        Close-up view of the controllable agent (blue) and the agent's corresponding trail $P_\alpha$(light blue) next to the target trail (red). The blue circle surrounding the agent represents the agent's vision divided into segments.}
    \end{center}
\end{figure}

We chose to reproduce the dynamics of the ant species \textit{Pristomyrmex punctatus} which is characterised by its lack of a queen caste, with mature colonies often consisting of thousands of female workers. Instead of constructing nests, the colony exhibits a strong clustering behaviour~\citep{Tsuji2011social}. (For further information on the experimental setup and the detection/tracking pipeline, see \cite{Maruyama2023}.)
Based on data extracted from video recordings, we built a simulation environment in which the agent (ant) begins each trial at position $P_\alpha(0) = P_\tau(0)$, the beginning of a (real) target ant trail.
\Cref{diag:arena-sim} shows the comparison between the video recordings and the simulation environment.
Currently, we select a target ant at random from the data whilst ensuring that the target moved a modest amount during the $t_\text{lim} = 30$ second time period. The target ant and corresponding trail are depicted in red, while the agent and agent trail are shown in blue. This is seen more clearly in \Cref{diag:ant-vision} where we also depict the ant's vision which we further separate into eight segments: $5$ in the forward-facing direction (indicated by the blue line depicting the agent's direction $\theta$), and $3$ facing backwards. Each ant in the environment begins at some position translated to pixel coordinates $(x, y)$ from the real data, relative to the arena's dimensions and scale.

Given two paths, the agent's path $P_\alpha(x)$ and the target path $P_\tau(x)$, the reward function $\mathcal{R}$ is defined as the scaled negative area between these paths over time, given by:
\begin{equation}
    R = -\sum_{t=1}^{t_\text{lim}} \left(1 - \frac{A_t}{\sqrt{1 + A_t^2}}\right)
\end{equation}
where $A_t$ represents the area of the trapezoid formed between the points along both trails at time $t$ and $t-1$.
Time $t$ represents a discrete time step in our simulation and reflects the real-time data underlying the ant movements. Each trial runs while $t \leq t_\text{lim}$.
This reward function is designed such that perfect alignment of the two trails yields a reward of $0$, with the reward becoming increasingly negative as their deviation increases.
However, alternative reward functions ought to be designed for more effective learning.

There are currently $13$ inputs for the agent observations: $\{x, y, s, \theta, \dot{\theta}, V_{fl_1}, V_{fl_2}, V_{fc}, V_{fr_2}, V_{fr_1}, V_r, V_b, V_l \}$ denoting the agent's position, speed, angle, rotation speed and visual observations. Each visual input provides a proportional measure of the number of ants detectable within the range and angle of vision. The outputs of the policy network are simply: \textsc{forward}, \textsc{backward}, \textsc{turn-left}, \textsc{turn-right}. (Due to our inability to detect pheromone trails from video recordings, these have been omitted.)

\section{Encoding Collective Behaviours}
\label{sec:neuroevo}
We intend to use this environment for the neuroevolution of models which accurately capture the natural dynamics of ant colonies.
The process of neuroevolution allows us to investigate how collective behaviours (both existing and novel) might be evolved through a series of gradual changes to the neural network that controls an individual such that, when replicated in a collective of individuals, intelligent collective behaviour emerges from their interactions.
More specifically, rather than focusing on adapting the weights of a network for a given task, we intend to focus on constructing the network topology itself, with the goal of encoding the task into the structure of the network to make inferring such a model more efficient, such as in the works of \cite{Gaier2019} and \cite{Stanley2002-NEAT}.
An alternative approach to growing neural networks for specific tasks was recently introduced by \cite{Najarro2023}.
Our purpose in evolving such policy networks is to understand the role that interaction bottlenecks within and between individuals play in the formation of their collective behaviours~\citep{Crosscombe2023-wh}, so that we might better develop novel collective behaviours in the future. 

Our environment is designed to not only replicate the intricate social interactions among ants but also to understand the emergent properties of their collective intelligence. By providing a simulation environment that captures real-time observational data from the target domain, researchers can iteratively evolve neural network configurations, promoting the emergence of sophisticated behavioural patterns akin to those found in nature.

\section{Conclusions}
\label{sec:conclusions}
We have introduced a new Gymnasium-based environment which features the reproduction of real ant colony behaviours that can be used to study the facets of emergent collective behaviour.
Eventually, we aim to encode a target collective behaviour into neural networks through a process of neuroevolution, with emphasis on the evolved topology as introduced by \cite{Gaier2019}. The evolved network can then be used as a policy network to control multiple agents which exhibit realistic collective behaviours inline with those observed in colonies of \textit{Pristomyrmex punctatus}. We hope that this work contributes to research into the necessary conditions for such collective behaviours to emerge.

\section{Acknowledgements}
Michael Crosscombe is supported by a Grant-in-Aid for JSPS Fellows 23KF0108. 
This work was partially supported by JSPS KAKENHI grants 21H04885 and 24H00707.

\footnotesize
\bibliographystyle{apalike}
\bibliography{references} 

\end{document}